\documentclass[aps,prd,preprint,showpacs,floats,nofootinbib]{revtex4}
\usepackage[dvips]{graphicx}
\usepackage{bm}
\usepackage{euscript,amsmath}

\begin{document}
\title{\boldmath{$B_d \to \pi^- K^{(*)+}$ and $B_s \to \pi^+(\rho^+) K^-$}  {\Large \bf decays with QCD factorization and flavor symmetry }}
\author {Guohuai Zhu}
\email[E-mail address: ]{zhugh@zju.edu.cn}
\affiliation{Zhejiang Institute of Modern Physics, Department of Physics, \\
 Zhejiang University, Hangzhou, Zhejiang 310027, P.R. China}

\date{\today}
\begin{abstract}
The QCD factorization (QCDF) method usually contains infrared
divergences which introduce large model dependence to its
predictions on charmless B decays. The amplitudes of charmless B
decays can be decomposed into ``tree" and ``penguin" parts which are
conventionally defined, not from the topology of the dominant
diagrams, but through their associated CKM factors $V_{ub}^*V_{uq}$
and $V_{tb}^*V_{tq}$, respectively, with $q=d,s$. We find that for
$B_{d,s} \to \pi^\mp K^\pm$ decays, the ``tree" amplitude can be
well estimated in QCDF with small errors, as the endpoint
singularities have been canceled to a large extent. With this as the
only input from QCDF and combined with flavor symmetry, the
branching ratio of $B_s \to \pi^+ K^-$ is estimated to be
significantly larger than the CDF measurement. This contradiction
could be solved if the form factor $F^{B_s K}$ is smaller than the
light cone sum rules estimation or the ``tree" amplitude has been
over estimated in QCDF. The latter possibility could happen if
charming penguins are nonperturbative and not small, as argued in
soft collinear effective theory. To differentiate between these two
possibilities, we examine the similar $B_s \to \rho^+ K^-$ decay
with the same technique. It is found that a large part of the
uncertainties are canceled in the ratio ${\cal B}(B_s \to \rho^+
K^-)/{\cal B}(B_s \to \pi^+ K^-)$. In QCDF, it is predicted to be
$2.5 \pm 0.2$  which is independent on the form factor. However if
charming penguins are important, this ratio could be very different
from the QCDF prediction. Therefore the ratio of these two branching
ratios could be an interesting indicator of the role of charming
penguins in charmless B decays.
\end{abstract}
\maketitle

\section{Introduction}

 Charmless hadronic B decays provide various
possibilities for the determination of the CKM angles, but often the
efforts are hindered by the low energy QCD dynamics which may lead
to large, sometimes even uncontrolled, systematic uncertainties. To
proceed further, one may resort to either factorization methods or
flavor symmetries. However it is known that here factorization only
holds to the lowest order of power expansion in $1/m_b$, which means
there is no (known) model-independent way to estimate power
corrections in $1/m_b$ systematically. For flavor symmetries, in
general SU(3) flavor symmetry breaking effects may also introduce
large uncertainties, though isospin is a good symmetry at the few
percent level. Recently, Descotes-Genon {\it et al.}
\cite{Descotes-Genon:2006wc} observed that, by the combined use of
QCD factorization (QCDF) and flavor symmetries, it is possible to
obtain new sum rules for $B_{d,s} \to K K$ decays with better
control on systematic uncertainties. This, together with the first
observation of $B_s \to \pi^+ K^-$ by CDF Collaboration
\cite{Aaltonen:2008hg}, stimulates us to investigate $B_d \to \pi^-
K^{(*)+}$ and $B_s \to \pi^+(\rho^+) K^-$ decays in the spirit of
Ref. \cite{Descotes-Genon:2006wc}.

Factorization aims to separate the calculable short distance effects from the
nonperturbative long distance part. In addition, the nonperturbative structure can be simplified as form factors,
decay constants and light-cone distribution amplitudes (LCDAs) in hadronic B decays at leading power expansion in $1/m_b$.
For charmless B decays, there are three factorization approaches available in the market: QCDF \cite{Beneke:1999br,Beneke:2000ry,Beneke:2003zv},
the perturbative QCD method (PQCD) \cite{Keum:2000ph,Keum:2000wi,Lu:2000em} and soft collinear effective theory (SCET)
\cite{Bauer:2000yr,Bauer:2001cu,Bauer:2004tj}. We shall not discuss PQCD in this paper.
For QCDF, phenomenologically power suppressed corrections such as chirally enhanced power corrections and
annihilation topology have to be included to fit the experimental data.
Unfortunately these terms contain endpoint singularities which break factorization and can only be estimated
in a model-dependent way. In SCET, the chirally enhanced power corrections and weak annihilation diagrams
are calculable \cite{Jain:2007dy,Arnesen:2006vb}, while the diagrams with internal charm quark loops are claimed to be
nonperturbative. Consequently these charming penguin amplitudes\footnote{It was first proposed in
\cite{Colangelo:1989gi,Ciuchini:1997hb,Ciuchini:2001gv} that the charming penguins
may play an important role in charmless B decays.}
can only be determined by fitting to the data.
Generally as weak annihilation and charming penguins have the same topology, it is hard in practice to
tell which one is really important. In this paper we find that the ratio
${\cal B}(B_s \to \rho^+ K^-)/{\cal B}(B_s \to \pi^+ K^-)$ may provide some insight into this issue.

This paper is organized as follows. In the next section, we first
discuss $B_{d,s} \to \pi^\mp K^\pm$ decays with a minimal use of
QCDF combined with flavor symmetry. We then extend the discussions
to $B_d \to \pi^- K^{*+}$ and $B_s \to \rho^+ K^-$ decays in the
third section. Finally we conclude with a summary in section IV.

\section{$B_{d,s} \to \pi^\mp K^\pm$ decays}

Recently the CDF Collaboration reported new results on the branching
ratios and direct CP violation for $B^0_{d,s}$ decay channels
\cite{Aaltonen:2008hg}, among them is the first observation of $B_s
\to \pi^+ K^- $ decay:
\begin{align}
{\cal B}(B_s \to \pi^+ K^-)& =(5.0 \pm 0.7 \pm 0.8 )\times
10^{-6}~, \\
A_{CP}(B_s \to \pi^+ K^-)& = ~0.39 \pm 0.15 \pm 0.08~,
\end{align}
where the first errors are statistical and the second systematic.
It was first proposed in \cite{Gronau:2000md} to
determine the CKM angle $ \gamma$ from $B_{d,s} \to \pi^\mp K^\pm$
decays using the U-spin symmetry. It was further noticed in
\cite{Lipkin:2005pb} that here the flavor symmetry breaking effects
should be unusually small since the strong phases from final state
interactions are exactly the same due to the charge conjugation
symmetry of the final states. Therefore it could be a robust test of
the Standard Model vs New Physics to check the equality of the
direct CP asymmetries of these two decay channels. In this paper we
will instead focus on their branching ratios, together with ($B_d
\to \pi^- K^{*+}$, $B_s \to \rho^+ K^-$) decays, with a minimal use
of QCDF combined with flavor symmetry\footnote{A recent analysis
of $B_{d,s} \to K \pi$ decays using flavor symmetry can be found in
\cite{Chiang:2008vc}.}.

For $\pi K$ channels, the decay amplitudes can usually be expressed as
\begin{align}
  A(B_d \to \pi^- K^+)&=A_{\pi K}^d (V_{ub}^* V_{us} T^d + V_{cb}^* V_{cs} P^d)~, \\
  A(B_s \to \pi^+ K^-)&=A_{\pi K}^s (V_{ub}^* V_{ud} T^s + V_{cb}^* V_{cd} P^s)~,
\end{align}
with
\[ A_{\pi K}^d=\frac{G_F}{\sqrt{2}}(m_{B_d}^2-m_\pi^2)f_K F^{B\pi}~, \hspace*{1cm}
 A_{\pi K}^s=\frac{G_F}{\sqrt{2}}(m_{B_s}^2-m_K^2)f_\pi F^{B_sK}~. \]
$T^{d,s}$ and $P^{d,s}$, multiplied by a common factor $A_{\pi K}^{d,s}$, denote the ``tree" and ``penguin" amplitudes
accompanied by the CKM factors $V_{ub}^* V_{uq}$ and $V_{cb}^* V_{cq}$, respectively, with $q=d,s$. In QCDF \cite{Beneke:2003zv},
\begin{align}\label{eq:TP}
  T&=\alpha_1+\alpha_4^u+\alpha_{4,ew}^u+\beta_3^u-\frac{1}{2}\beta_{3,ew}^u~, \nonumber \\
  P&=\alpha_4^c+\alpha_{4,ew}^c+\beta_3^c-\frac{1}{2}\beta_{3,ew}^c~,
\end{align}
where the weak annihilation contributions are contained in $\beta$ terms, while all the other
parts are grouped into $\alpha$ terms, including the vertex corrections and the hard spectator scattering contributions.
The explicit expressions of $\alpha$'s and $\beta$'s can be found in \cite{Beneke:2003zv}.
In QCDF, the endpoint singularities appear in annihilation topology and hard spectator scattering diagrams,
which introduce significant uncertainties to $\alpha$'s and $\beta$'s. Therefore neither T nor P
can be predicted reliably in QCDF with small errors. However, by implementing the unitarity condition of the CKM matrix, the above decay
amplitudes can be reexpressed as
\begin{align}
  A(B_d \to \pi^- K^+)&=A_{\pi K}^d (V_{ub}^* V_{us} {\widetilde T}^d - V_{tb}^* V_{ts} P^d)~, \\
  A(B_s \to \pi^+ K^-)&=A_{\pi K}^s (V_{ub}^* V_{ud} {\widetilde T}^s - V_{tb}^* V_{td} P^s)~,
\end{align}
with ${\widetilde T}=T-P$. It was first observed in \cite{Descotes-Genon:2006wc} that for
$B_{d,s} \to K^0 \bar{K}^0$ decays, although both T and P contain endpoint singularities,
${\widetilde T}$ is infrared safe as the endpoint singularities in T and P cancel completely.
Therefore ${\widetilde T}$ can be estimated reliably in QCDF. However such case is rare in B decays.
For $B \to K\pi$ decays, using Eq. (\ref{eq:TP}), one obtains
\begin{equation}\label{eq: T expression}
  {\widetilde T}=\alpha_1+(\alpha_4^u-\alpha_4^c)+(\alpha_{4,ew}^u-\alpha_{4,ew}^c)+
  (\beta_3^u-\beta_3^c)-\frac{1}{2}(\beta_{3,ew}^u-\beta_{3,ew}^c)~.
\end{equation}
Interestingly, for every bracket in the above equation, the endpoint singularities are canceled exactly, so the only
residual endpoint singularity lies in $\alpha_1$ which comes from the hard spectator scattering diagrams. Therefore,
although ${\widetilde T}$ in $B \to K\pi$ decays is not totally free of infrared divergence, in practice it could
be estimated in QCDF with much smaller uncertainty.

Numerically, the errors of ${\widetilde T}$ are dominated by the
residual endpoint divergence from hard spectator scattering diagrams
which is usually modeled as
\begin{align}
    X_H=\ln \left ( \frac{m_B}{\Lambda_h}\right )\left ( 1+ \rho_H e^{i\phi_H}\right )~,
\end{align}
and the parameter $\lambda_B$, which parameterizes the integral of the B
meson wave function as
\begin{align}
   \int_0^1\frac{\Phi_1^B(\xi)}{1-\xi} d\xi \equiv \frac{m_B}{\lambda_B}~.
\end{align}
To estimate the uncertainties, we will take $\lambda_B=300 \pm 100$ MeV, $0 \leq \rho_H \leq 1$,
$0 \leq \phi_H \leq 2\pi$ and adopt the default values of \cite{Beneke:2003zv} for all the other parameters.
We then find
\begin{align}\label{eq: T result}
   \vert {\widetilde T}^s \vert=0.99^{+0.03}_{-0.05}~,\hspace*{2cm}  \vert {\widetilde T}^d \vert=1.00^{+0.03}_{-0.04}~,
\end{align}
which respects the U-spin symmetry ${\widetilde T}^s={\widetilde T}^d$ at one percent level.
Thus in the following we will not differentiate between ${\widetilde T}^s$ and ${\widetilde T}^d$ and simply take in
the flavor symmetry limit
\begin{align}
  \vert {\widetilde T}^d \vert \simeq \vert {\widetilde T}^s \vert \equiv {\widetilde T} = 0.99^{+0.03}_{-0.05}~.
\end{align}
It is worth reminding that charming penguins, accompanied by the CKM factors\ $V_{cb}^* V_{cq}$, will contribute
to both ${\widetilde T}$ and P. So if charming penguins are nonperturbative and not small as argued in SCET, the
above estimation of ${\widetilde T}$ could change.

As mentioned earlier, in QCDF the ``penguin" amplitude P is significantly affected by the endpoint divergences and
therefore hard to be estimated in a reliable way. So
the above equation of ${\widetilde T}$ is the only result that we will adopt from QCDF.
In this sense $B_{d,s} \to K\pi$ decays are analyzed with``a minimal use of QCDF". For all the rest
we will resort to flavor symmetry.
As observed in \cite{Lipkin:2005pb}, the flavor symmetry breaking effects in $B_{d,s} \to \pi^\mp K^\pm$ should be
unusually small due to charge conjugation which is a strict symmetry for the strong phases from final state interactions.
So by taking the factors $A_{\pi K}^{d,s}$ out which have included explicitly part of the U-spin symmetry breaking effects,
it should be reasonable to take the approximation of the U-spin symmetry limit $P^d \simeq P^s\equiv P$. This amounts to
neglect hard spectator scattering contributions sensitive to the difference between $B_d$ and $B_s$ distributions, and also
annihilation contributions when the gluon is emitted from the spectator quark. Then the decay amplitudes can be rewritten as
\begin{align}
  A(B_d \to \pi^- K^+)&=A_{\pi K}^d (V_u e^{i\gamma} \vert {\widetilde T} \vert - V_t \vert P \vert e^{i\delta})~, \\
  A(B_s \to \pi^+ K^-)&=A_{\pi K}^s (V'_u e^{i\gamma} \vert{\widetilde T}\vert + V'_t e^{-i\beta} \vert P \vert e^{i\delta})~,
\end{align}
where $V_q \equiv \vert V_{qb}^* V_{qs} \vert$ and $V'_q \equiv \vert V_{qb}^* V_{qd} \vert$ for $q=u,~t$.
$\delta$ is the relative strong phase between P and ${\widetilde T}$:
\[ P/{\widetilde T}=-\vert P/{\widetilde T} \vert e^{i\delta}~, \]
where the convention is chosen such that $\delta=0$ in the naive factorization limit.
It is then straightforward to get a well-known relation
between the direct CP violations of $B_{d,s} \to \pi K$ decays
\begin{align}
 \frac{A_{CP}(B_s \to \pi^+ K^-)}{A_{CP}(B^0 \to \pi^- K^+)} =
-\frac{{\cal B}(B^0 \to \pi^- K^+)}{{\cal B}(B_s \to \pi^+ K^-)}\frac{\tau(B_s)(A^s_{\pi K})^2}{\tau(B^0)(A^d_{\pi K})^2}~.
\end{align}
Concerning the currently large experimental uncertainty of $A_{CP}(B_s \to \pi^+ K^-)$,
this relation is well consistent within errors with the current data of $B_s \to \pi K$ in Eqs.(1,2) and of $B_d \to \pi K$ \cite{Barberio:2008fa}
\begin{align}\label{eq:Bd observable}
{\cal B}(B_d \to \pi^- K^+)=(19.4 \pm 0.6) \times 10^{-6}~,\hspace*{1.cm} A_{CP}(B_d \to \pi^- K^+)=(-9.8^{+1.2}_{-1.1})\%~.
\end{align}
In this sense, there are three independent experimental observables:
two branching ratios and one direct CP violation. If the related
form factors are taken as input, there are also three unknowns: CKM
angle $\gamma$, the ``penguin" amplitude $\vert P \vert$, and the
strong phase $\delta$. Then in principle, it should be possible to
extract $\gamma$ from $B_{d,s} \to \pi^\mp K^\pm$ decays, if the
QCDF knowledge of ${\widetilde T}$ is reliable. However in practice,
it is hard to obtain a strong constraint on the CKM angle $\gamma$
through this way, especially considering the uncertainties of the
input parameters and also the mild dependence of the branching
ratios on the angle $\gamma$. As currently no clear evidence of new
physics beyond SM exists in rare B decays, we will instead take the
global fit determination of $\gamma=(67.9^{+4.3}_{-3.8})^\circ$ from
CKMfitter \cite{Charles:2004jd} as input. Then our focus is what
information one may obtain about the soft QCD dynamics in $B \to
K\pi$ decays.

Now only the ``penguin" amplitude $\vert P \vert$ and the strong phase $\delta$ are unknown, which can be
extracted solely from the observables of $B_d \to \pi^- K^+$ decay in Eq. (\ref{eq:Bd observable}).
With the CKM parameters taken from CKMfitter \cite{Charles:2004jd}:
\begin{align}\label{eq: CKMfitter input}
   \lambda=0.2251~, \hspace*{0.5cm} \vert V_{ts} \vert=0.04042^{+0.00037}_{-0.00118}~,\hspace*{0.5cm}
   \vert V_{ub} \vert=0.00351^{+0.00015}_{-0.00016}~,
\end{align}, the determination of $\vert V_{ub} \vert F^{B\pi}=(9.1 \pm
0.7)\times 10^{-4}$ \cite{Ball:2006jz} from the semi-leptonic $B \to
\pi \ell \bar{\nu}$ decay and $\tau_{B_d}=1.525 \pm 0.009$~ps
\cite{Barberio:2008fa}, we obtain
\begin{align}\label{eq: P result}
\vert P \vert = 0.129 \pm 0.012~, \hspace*{1cm} \delta=-(18.9\pm 2.9)^\circ~,
\end{align}
with the uncertainties coming mainly from the variation of $\vert
V_{ub} \vert F^{B\pi}$. Notice that the determination of $\vert P
\vert$ is rather insensitive to the change of ${\widetilde T}$. For
instance, about $20\%$ change of $\vert {\widetilde T} \vert$ from
$0.99$ to $0.80$, with all the other parameters unchanged, will lead
to only $1\%$ variation of the central value of $\vert P \vert$ from
$0.129$ to $0.128$.
This is because, as a penguin dominant decay, the branching ratio of
$B_d \to \pi^- K^+$ decay largely determines the magnitude of $P$,
irrespective of how much the charming penguins may contribute to
$P$.

It is also known that $B^+ \to \pi^+ K^0$ decay is almost pure
penguin process, in addition its penguin amplitude is equal to that
of $B_d \to \pi^- K^+$ decay except for the negligible
color-suppressed electroweak penguin contributions. As a consistency
check, we do find that the penguin amplitude of $B^+ \to \pi^+ K^0$
is $\vert P \vert=0.130 \pm 0.012$, which is in well agreement with
Eq. (\ref{eq: P result}). Therefore the above estimation of the
``penguin" amplitude $\vert P \vert$ in Eq. (\ref{eq: P result})
should be, up to few percent, independent on whether the charming
penguins are important or not.

\begin{table}
\begin{tabular}{|c|c|c|c|c|}\hline
 $10^2 \vert V_{ts}\vert$ & $f_\pi$ & ~~$10^4 \vert V_{ub}\vert F^{B\pi}$~~ & $\tau_{B_s}$ & ~~$10^6{\cal B}(B^+ \to \pi^+ K^0)$~~\\ \hline
 ~~$4.042^{+0.037}_{-0.118}~~$ &  ~~$131$ MeV~~ & ~~$9.1 \pm 0.7$~~ & ~~$1.472^{+0.024}_{-0.026}$~ps~~ &$23.1 \pm 1.0$\\ \hline
 $10^3 \vert V_{td}\vert$ & $f_K$ &$F^{B_sK}/F^{B\pi}$ &$\tau_{B^+}$ &   \\ \hline
 ~~$8.59^{+0.28}_{-0.29}$~~&~~$160$ MeV~~ &~~$1.15^{+0.17}_{-0.09}$~~&~~$1.638 \pm 0.011$~ps~~ &   \\ \hline
\end{tabular}
\caption{Input parameters in numerical calculations of $B_s \to
\pi^+ K^-$ decay. The CKM parameters are taken from CKMfitter
\cite{Charles:2004jd}, The $B$-meson lifetimes and the branching
ratio are taken from \cite{Barberio:2008fa} and the form factors are
given in \cite{Ball:2006jz,Duplancic:2008tk}. }
\end{table}

Taking Eq. (\ref{eq: P result}) as input, we can now estimate the
branching ratio of $B_s \to \pi^+ K^-$ decay. The CP-averaged
amplitude square can be expressed as
\begin{align}
\vert A(B_s \to \pi^+ K^-) \vert^2=( A_{\pi K}^s)^2  \vert
V'_u{\widetilde T} \vert^2 \left ( 1 +\left \vert \frac{V'_t}{V'_u}
\frac{P}{{\widetilde T}} \right \vert^2 +2 \left \vert
\frac{V'_t}{V'_u} \frac{P}{{\widetilde T}}\right \vert \cos \delta
\cos (\beta+\gamma)\right )~.
\end{align}
Due to the accidental fact that $\beta+\gamma \sim 90^\circ$, the
``tree"-``penguin" interference term in $B_s \to \pi^+ K^-$ decay is
negligibly small. Thus approximately the strong phase $\delta$ does
not affect the branching ratio and one has
\begin{align}
\vert A(B_s \to \pi^+ K^-) \vert^2 \simeq (A_{\pi K}^s)^2 \left (
\vert V'_u {\widetilde T} \vert^2 + \vert V'_t P \vert^2 \right )~.
\end{align}
If one observes that the $\vert P \vert^2$ term  can be directly related
to the pure penguin process $B^+ \to \pi^+ K^0$ as
\begin{align}\label{eq: P relation}
\left (\frac{V'_t}{V_t} \frac{A_{\pi K}^s}{A_{\pi K}^d} \right )^2
\vert A(B^+ \to \pi^+ K^0) \vert^2~,
\end{align}
numerically one can then obtain
\begin{align}\label{eq: piK final}
{\cal B}(B_s \to \pi^+ K^-) = \left (\frac{F^{B_sK}}{F^{B\pi}}
\right )^2 \left [ (6.2 \pm 1.0)\vert {\widetilde T} \vert^2 + 0.6
\right ] \times 10^{-6}~,
\end{align}
with the input parameter listed in Table I. Here the uncertainty of
$\vert P \vert$ term, which can be estimated by using Eq. (\ref{eq:
P relation}), is much smaller than that of ${\widetilde T}$ term and
has been therefore neglected. We want to stress that Eq. (\ref{eq:
piK final}) is not just valid in QCDF, but rather be a model
independent result to a large extent.

The form factors ratio has been estimated in light-cone sum rules
\cite{Duplancic:2008tk}, as shown in Table I. As discussed earlier,
${\widetilde T}$ can also be estimated in a relatively reliable way
in QCDF in Eq. (\ref{eq: T result}), then we get \footnote{A direct
QCDF calculation in \cite{Beneke:2003zv} gives
$10.2^{+4.5+3.8+0.7+0.8}_{-3.9-3.2-1.2-0.7} \times 10^{-6}$ as their
``default results". Here the first error corresponds to the
variations of the CKM factors and the second error comes mainly from
the variation of the form factor. The readers are referred to
\cite{Beneke:2003zv} for more details of the analysis.}
\begin{align}
{\cal B}^{th}(B_s \to \pi^+ K^-)=8.8^{+1.4+2.8}_{-1.4-1.3} \times
10^{-6}~,
\end{align}
where the first error comes mainly from the uncertainties of $\vert
V_{ub} \vert F^{B\pi}$ and ${\widetilde T}$, while the second error
comes from the form factors ratio. This estimation is significantly
larger than the CDF measurement $(5.0 \pm 1.1) \times 10^{-6}$.

Notice that $\vert V_{ub} \vert F^{B\pi}$ is extracted from $B \to
\pi \ell \nu$ decays in a model independent way. Thus if the updated
experimental measurement does not change greatly, it must be that
either the form factors ratio is much smaller than the light-cone
sum rules prediction or ${\widetilde T}$ is overestimated in QCDF,
or both. The possibility of a smaller ratio of the form factors has
been discussed in a flavor symmetry analysis \cite{Chiang:2008vc}
and also in a recent QCDF analysis \cite{Cheng:2009mu,Cheng:2009cn}
where the central values of $F^{B_s K}=0.24$ and $F^{B\pi}=0.25$ are
taken. The latter possibility of a smaller ${\widetilde T}$ can not
be easily realized in QCDF, though it could happen in SCET if
charming penguins are not small. In practice, however, it is
difficult to tell experimentally which possibility is chosen by
nature. This is because ${\widetilde T}$ is not directly observable
in charmless B decays. For the form factor $F^{B_s K}$, in principle
it could be determined via semileptonic $B_s \to K \ell \nu$ decays.
But this measurement is not easy at hadron colliders, such as the
Tevatron and the LHC. For the future super B factories, the
challenge is to accumulate enough $B_s$ mesons for precise
measurement of the kaon momentum spectrum of this semileptonic
decay. $F^{B_s K}$ may also be predicted by lattice QCD computations
at small recoil, while unfortunately charmless B decays happen at
large recoil.

\section{$B_d \to \pi^- K^{*+}$ and $B_s \to \rho^+ K^-$ decays}

To find a way out of this problem, we may examine similar decay
channels $B_d \to \pi^- K^{*+}$, $B_s \to \rho^+ K^-$ and $B^+ \to
\pi^+ K^{*0}$. At first sight, it may seem better to discuss
$B_{d,s} \to \rho^\mp K^\pm$ decays as the final states are related
by charge conjugation symmetry. However these two decay channels are
actually not related by U-spin symmetry. Again we may express the
relevant decay amplitudes in the following form:
\begin{align}
  A(B_d \to \pi^- K^{*+})&=A_{\pi K^*} (V_u e^{i\gamma} \vert {\widetilde T_{K^*}} \vert - V_t \vert P_{K^*} \vert e^{i\delta_{K^*}})~, \\
  A(B_s \to \rho^+ K^-)&=A_{\rho K} (V'_u e^{i\gamma} \vert{\widetilde T_\rho}\vert + V'_t e^{-i\beta} \vert P_\rho \vert e^{i\delta_\rho})~,
\end{align}
with
\[ A_{\pi K^*}=\sqrt{2}G_F m_{B_d} p_c f_{K^*} F^{B\pi}~, \hspace*{1cm}
 A_{\rho K}=\sqrt{2}G_F m_{B_s} p_c f_\rho F^{B_sK}~. \]
Here $p_c$ is the c.m. momentum of the final state mesons. In QCDF,
the expression of ${\widetilde T_\rho}$ is formally the same as Eq.
(\ref{eq: T expression}) of ${\widetilde T}$, though the explicit
formulae of the coefficients $\alpha$'s and $\beta$'s are different
which can also be found in \cite{Beneke:2003zv}. Since now the final
states $\pi K^*$ and $\rho K$ are not related to each other under
charge conjugation, there is no reason to expect that here the
U-spin symmetry breaking effects would be particularly small. This
means the difference between (${\widetilde T}_\rho$, $P_\rho$,
$\delta_\rho$) and (${\widetilde T}_{K^*}$, $P_{K^*}$,
$\delta_{K^*}$) could reach the level of $20\%$ and can not be
simply ignored. But if we focus only on the $B_s \to \rho^+ K^-$
decay, the strong phase $\delta_\rho$ again does not affect the
CP-averaged branching ratio
\begin{align}
\vert A(B_s \to \rho^+ K^-) \vert^2 \simeq (A_{\rho K}^s)^2 \left (
\vert V'_u {\widetilde T}_\rho \vert^2 + \vert V'_t P_\rho \vert^2
\right )~
\end{align}
in good approximation due to the accidental fact that $\beta+\gamma
\simeq 90^\circ$, just like the case of $B_s \to \pi^+ K^-$ decays.
Similarly we find $\vert {\widetilde T}_\rho
\vert=0.99^{+0.03}_{-0.05}$ with the uncertainties mainly from
$\lambda_B$ and the hard spectator parameters $\rho_H^{PV}$ and
$\phi_H^{PV}$. We will soon find the ratio ${\widetilde
T}_\rho/{\widetilde T}$ to be useful. As they share the common
parameter $\lambda_B$, their errors are partly correlated and the
uncertainty related to $\lambda_B$ will be canceled in the ratio.
But the hard spectator parameters are independent for $B_s \to PP$
and $B_s \to PV$ decays. Concerning this, we find the ratio to be
\[ \left \vert \frac{\widetilde T_\rho}{\widetilde T} \right \vert=1.00 \pm 0.04  \]
with the uncertainty coming dominantly from the variation of hard
spectator parameters.

Then the last missing piece required is the ``penguin" amplitude
$\vert P_\rho \vert$. Notice that $B_s \to \rho^+ K^-$ decays are
dominated by the ``tree" amplitude ${\widetilde T}_\rho$.
Furthermore it is known that generally the penguin amplitudes of $B
\to PV$ decays are smaller than those of $B \to PP$ decays. As the
$\vert P \vert$ term contribute about $10\%$ of the final branching
ratio of $B_s \to \pi^+ K^-$ decay, it is expected that the
contribution of $\vert P_\rho \vert$ term should be even smaller.
Therefore for such a small term, an estimation based on SU(3) flavor
symmetry should be enough which will not introduce large uncertainty
to the final branching ratio result.

In the flavor symmetry limit, one has $\vert P_\rho \vert=\vert
P_{K^*} \vert$. Then one may estimate $\vert P_{K^*} \vert$ from
either $B_d \to \pi^- K^{*+}$ decay or the almost pure penguin $B^+
\to \pi^+ K^{*0}$ decay, just like the case of $B \to \pi K$ decays.
Taking $f_\rho=216$ MeV, $f_{K^*}=220$ MeV and the parameters listed
in Table I, we find $\vert P_{K^*} \vert=0.066^{+0.008}_{-0.006}$
using \cite{Barberio:2008fa}
\begin{align}\label{eq:piks observable}
{\cal B}(B_d \to \pi^- K^{*+})=(8.6 \pm 0.9) \times
10^{-6}~,\hspace*{1.cm} A_{CP}(B_d \to \pi^- K^{*+})=(-18 \pm 7)\%~,
\end{align}
which is consistent with the determination of $\vert P_{K^*}
\vert=0.064^{+0.007}_{-0.006}$ from ${\cal B}(B^+ \to \pi^+
K^{*0})=(9.9^{+0.8}_{-0.9}) \times 10^{-6}$ \cite{Barberio:2008fa}.
Then we obtain
\begin{align}
\vert A(B_s \to \rho^+ K^-) \vert^2 \simeq (A_{\rho K}^s)^2 \left (
\vert V'_u {\widetilde T}_\rho \vert^2 + \left (\frac{V'_t}{V_t}
\frac{\vert A(B^+ \to \pi^+ K^{*0}) \vert}{A_{\pi K^*}} \right )^2
\right )~.
\end{align}
Numerically the corresponding branching ratio can then be estimated
\begin{align}\label{eq: rhoK final}
{\cal B}(B_s \to \rho^+ K^-) = \left (\frac{F^{B_sK}}{F^{B\pi}}
\right )^2 \left [ (15.7 \pm 2.4)\vert {\widetilde T}_\rho \vert^2 +
0.4 \right ] \times 10^{-6}~.
\end{align}
where the uncertainty from $\vert P_\rho \vert$ term is much smaller
than that from ${\widetilde T}_\rho$ term and has therefore been
ignored. We want to stress again that Eq. (\ref{eq: rhoK final}),
just like Eq. (\ref{eq: piK final}), is a model independent result
to a large extent.

Notice that $B_s \to \pi^+ (\rho^+) K^-$ decays depend upon many
common parameters such as $F^{B_s K}$, $\vert V_{ub} \vert F^{B\pi}$
etc., their uncertainties are highly correlated. It is then clear
that the ratio of the branching ratios should have much smaller
error. Actually the error of the ratio depends almost solely on
${\widetilde T}_\rho/{\widetilde T}$
\begin{align}
\frac{{\cal B}(B_s \to \rho^+ K^-)}{{\cal B}(B_s \to \pi^+ K^-)}
\simeq \frac{f_\rho^2}{f_\pi^2} \left \vert \frac{{\widetilde
T}_\rho}{\widetilde T} \right \vert^2 \frac{1+0.4/(15.7 \vert
{\widetilde T}_\rho \vert^2) }{1+0.6/(6.2\vert {\widetilde T}
\vert^2)} \simeq 2.5 \left \vert \frac{{\widetilde
T}_\rho}{\widetilde T} \right \vert^2~.
\end{align}
In the above estimation, we have adopted flavor symmetries to relate
the ``penguin" amplitude of $B_s \to \pi^+ (\rho^+) K^-$ to that of
$B_d \to \pi^- K^{(*)+}$, which may introduce $20\%$ level of
uncertainty. But fortunately as tree-dominant decays, the ratio
${\cal B}(B_s \to \rho^+ K^-)/{\cal B}(B_s \to \pi^+ K^-)$ is not
sensitive to the variation of the relevant ``penguin" amplitudes. As
an illustration, we assign conservatively $30\%$ uncertainty to the
``penguin" amplitudes to take account of the flavor symmetry
breaking effects, and adopt the QCDF estimation $\vert {\widetilde
T_\rho}/{\widetilde T} \vert = 1.00 \pm 0.04$, it is found that
\begin{align}\label{eq: ratio}
\left . \frac{{\cal B}(B_s \to \rho^+ K^-)}{{\cal B}(B_s \to \pi^+
K^-)} \right \vert_{QCDF} \simeq 2.5 \pm 0.1 \pm 0.2~,
\end{align}
with the first error comes from the flavor symmetry breaking effects
and the second error comes from the variation of $\vert {\widetilde
T_\rho}/{\widetilde T} \vert$.
 While in SCET, if the charming penguins are not small, it could
alter ${\widetilde T}$ and ${\widetilde T}_\rho$. It is worth
reminding that there is no known reason, at least to our knowledge,
to expect that the charming penguins of $B_s \to \pi^+ K^-$ and
$\rho^+ K^-$ channels should be the same. This means the ratio
${\cal B}(B_s \to \rho^+ K^-)/{\cal B}(B_s \to \pi^+ K^-)$ in SCET
could be quite different from Eq. (\ref{eq: ratio}). For example, a
recent global fit analysis of charmless B decays using SCET gives
\cite{Wang:2008rk}\footnote{The results of $\pi^+ K^-$ were provided
by Wei Wang, one author of \cite{Wang:2008rk}, in a private
communication.} (in units of $10^{-6}$)
\begin{align}
   {\cal B}(B_s \to \pi^+ K^-)&=5.7^{+0.6+0.5}_{-0.6-0.5}~, \hspace*{0.5cm}
   {\cal B}(B_s \to \rho^+ K^-)=7.6^{+0.3+0.8}_{-0.1-0.8} \hspace*{0.7cm} \mbox{(Solution I)} \\
   {\cal B}(B_s \to \pi^+ K^-)&=5.5^{+0.6+0.5}_{-0.6-0.4}~, \hspace*{0.5cm}
   {\cal B}(B_s \to \rho^+ K^-)=10.2^{+0.4+0.9}_{-0.5-0.9} \hspace*{0.5cm} \mbox{(Solution II)}
\end{align}
In these two solutions, the ratio ${\cal B}(B_s \to \rho^+
K^-)/{\cal B}(B_s \to \pi^+ K^-)$ equals to $1.3$ or $1.9$,
significantly different from the QCDF prediction $2.5 \pm 0.2$.
Therefore a precise measurement of this ratio in the near future
could be an interesting test of the importance of the charming
penguins in charmless B decays.

\section{Summary}

In this paper we first studied $B_{d,s} \to \pi^\mp K^\pm$ decays
using QCDF combined with flavor symmetry. In general, the QCDF
calculations contain infrared divergence when weak annihilation
topology and hard spectator scattering diagrams are included. But if
we express the decay amplitude as
\[ A(B_s \to \pi^+ K^-)=A_{\pi K}^s (V_{ub}^* V_{ud} {\widetilde T} - V_{tb}^* V_{td} P)~, \]
The ``tree" amplitude ${\widetilde T}$ can be (relatively) reliably
estimated in QCDF to be $0.99^{+0.03}_{-0.05}$. This is because the
endpoint singularities inside ${\widetilde T}$ has been canceled to
a large extent. In addition, the CP-averaged branching ratio is insensitive to
the relative strong phase between ${\widetilde T}$ and $P$ because the interference term
is proportional to $\cos (\beta+\gamma)$ which is (accidently) very close to zero.
Noticed that the final states of $B_{d,s} \to \pi^\mp K^\pm$ are invariant under charge conjugation, implying that the flavor symmetry
breaking effect should be unusually small as pointed out in \cite{Lipkin:2005pb}. So
the penguin amplitude $\vert P \vert$ can be determined from $B_d \to \pi^- K^+$ decay with small error.
Then we obtain
\begin{align}
{\cal B}(B_s \to \pi^+ K^-) = \left (\frac{F^{B_sK}}{F^{B\pi}}
\right )^2 \left [ (6.2 \pm 1.0)\vert {\widetilde T} \vert^2 + 0.6
\right ] \times 10^{-6}~.
\end{align}
With the light cone sum rules estimation \cite{Duplancic:2008tk}
$F^{B_sK}/F^{B\pi}=1.15^{+0.17}_{-0.09}$ and the QCDF prediction of
${\widetilde T}$, the expected branching ratio would be
$8.8^{+3.1}_{-1.9} \times 10^{-6}$, much larger than the CDF
observation $(5.0 \pm 1.1) \times 10^{-6}$. One possibility is that
the form factors ratio $F^{B_sK}/F^{B\pi}$ has been overestimated by
the light cone sum rules calculation. But it is also possible that
the QCDF prediction of ${\widetilde T}$ is too large. This could
happen, for example, if the charming penguins play an important role
in charmless B decays.

To differentiate between these two possibilities, we examined the
similar decay channels $B_s \to \rho^+ K^-$, $B_d \to \pi^- K^{*+}$
and $B^+ \to \pi^+ K^{*0}$ with the same method. We find a large
part of uncertainties has been canceled in the ratio ${\cal B}(B_s
\to \rho^+ K^-)/{\cal B}(B_s \to \pi^+ K^-)$ which equals
approximately to be $2.5 \vert {\widetilde T}_\rho/{\widetilde T}
\vert^2$. In QCDF, $\vert {\widetilde T}_\rho/{\widetilde T} \vert$
is also well predicted to be about one, so we get an interesting
prediction ${\cal B}(B_s \to \rho^+ K^-)/{\cal B}(B_s \to \pi^+
K^-)=2.5 \pm 0.1 \pm 0.2$ in QCDF, with the first error reflects the
flavor symmetry breaking effects and the second error reflects the
uncertainty of $\vert {\widetilde T}_\rho/{\widetilde T} \vert$.
However if charming penguins are not small, this ratio could vary
significantly from $2.5$, which means the ratio of these two
branching ratios could be useful to clarify the role of charming
penguins in charmless B decays.

\section*{Acknowledgement}

This work is supported in part by the National Science Foundation of
China under grant No. 10705024 . G.Z is also supported in part by
Chinese Universities Scientific Fund and in part by the Scientific
Research Foundation for the Returned Overseas Chinese Scholars,
State Education Ministry. G.Z thanks the Kavli Institute for
Theoretical Physics China (KITPC) for their hospitality where part
of this work was completed.

\end{document}